\documentclass[%
 preprint,
amsmath,amssymb,
prl,
longbibliography,
superscriptaddress,
]{revtex4-1}

\usepackage{graphicx}% Include figure files
\usepackage{dcolumn}% Align table columns on decimal point
\usepackage{bm}% bold math
\usepackage{amsmath}
\usepackage{color}
\usepackage{float}

\begin{document}

\title{Quasi-edge states and topological Bloch oscillation
\\ in the synthetic space}

\author{Xiaoxiong Wu}
\affiliation{State Key Laboratory of Advanced Optical Communication Systems and Networks, School of Physics and Astronomy, Shanghai Jiao Tong University, Shanghai 200240, China}

\author{Luojia Wang}
\affiliation{State Key Laboratory of Advanced Optical Communication Systems and Networks, School of Physics and Astronomy, Shanghai Jiao Tong University, Shanghai 200240, China}

\author{Guangzhen Li}
\affiliation{State Key Laboratory of Advanced Optical Communication Systems and Networks, School of Physics and Astronomy, Shanghai Jiao Tong University, Shanghai 200240, China}

\author{Dali Cheng}
\affiliation{State Key Laboratory of Advanced Optical Communication Systems and Networks, School of Physics and Astronomy, Shanghai Jiao Tong University, Shanghai 200240, China}

\author{Danying Yu}
\affiliation{State Key Laboratory of Advanced Optical Communication Systems and Networks, School of Physics and Astronomy, Shanghai Jiao Tong University, Shanghai 200240, China}

\author{\\ Yuanlin Zheng}
\affiliation{State Key Laboratory of Advanced Optical Communication Systems and Networks, School of Physics and Astronomy, Shanghai Jiao Tong University, Shanghai 200240, China}

\author{Vladislav V. Yakovlev}
\affiliation{Texas A{\upshape{\&}}M University, College Station, Texas 77843, USA}

\author{Luqi Yuan}
\email{yuanluqi@sjtu.edu.cn}
\affiliation{State Key Laboratory of Advanced Optical Communication Systems and Networks, School of Physics and Astronomy, Shanghai Jiao Tong University, Shanghai 200240, China}

\author{Xianfeng Chen}
\affiliation{State Key Laboratory of Advanced Optical Communication Systems and Networks, School of Physics and Astronomy, Shanghai Jiao Tong University, Shanghai 200240, China}
\affiliation{Shanghai Research Center for Quantum Sciences, Shanghai 201315, China}
\affiliation{Jinan Institute of Quantum Technology, Jinan 250101, China}
\affiliation{Collaborative Innovation Center of Light Manipulation and Applications, Shandong Normal University, Jinan 250358, China}

%\date{\today}

\begin{abstract}
In physics, synthetic dimensions trigger great interest to manipulate light in different ways, while in technology, lithium niobate shows important capability towards on-chip applications. Here, based on the state-of-art technology, we propose and study a theoretical model of dynamically-modulated waveguide arrays with the Su-Schrieffer-Heeger configuration in the spatial dimension. The propagation of light through the one-dimensional waveguide arrays mimics time evolution of field in a synthetic two-dimensional lattice including the frequency dimension. By adding the effective gauge potential, we find \textit{quasi-edge state} that the intensity distribution manifests not at the boundary as the traditional edge state, which leads to an exotic topologically protected one-way transmission along adjacent boundary. Furthermore, a cosine-shape isolated band exhibits, supporting the topological Bloch oscillation in the frequency dimension under the effective constant force, which is localized at the spatial boundary and shows the topological feature. Our work therefore points out further capability of light transmission under topological protections in both spatial and spectral regimes, and provides future on-chip applications in the lithium niobate platform.
\end{abstract}

\maketitle

%\tableofcontents
Synthetic dimensions in photonics have been under extensive explorations recently in both
theory and experiment \cite{yuan2018optica,ozawa2019,yuan2021apl,lustig2021}.  Among versatile studies using different degrees of freedom of light \cite{degree1,degree2,degree3,degree5yuan2016ol,degree4,degree6,degree8,
degree9yuan2019prl,frequency1dutt2019nc,degree10,degree11}, the optical frequency has been pointed out to be a good candidate along which synthetic dimension can be constructed \cite{yuan2018optica}. In particular, both dynamically-modulated ring resonator system and modulated nonlinear waveguide have been shown as potential platforms to explore different physical phenomena in a synthetic space including the frequency axis of light in experiments \cite{degree7qin2018prl,frequency2dutt2020science,frequency3li2021sci}. A variety of physical phenomena such as one-dimensional Bloch oscillation \cite{bo1yuan2016optica,bo2qin2018pra,bo3}, topological insulators \cite{degree8,frequency2dutt2020science}, non-Hermitian  Hamiltonian \cite{frequency4,nonher}, and higher-order topological physics \cite{high1dutt,high2} have been discussed with the synthetic frequency dimension previously.

\begin{figure}[htb]
\center
\includegraphics[width=16cm]{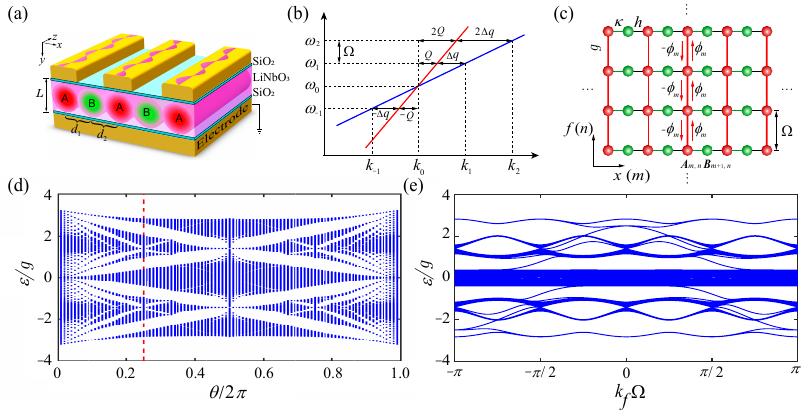}
\caption{\label{fig1} (a) The schematic of waveguide arrays based on LNOI technology, where waveguides labeled as A and B alternatively arranged in-between SiO$_2$ buffering layers and only waveguides of type A are modulated by external voltages. (b) Travelling-wave dynamic modulation (marked as red line) and intrinsic dispersion curve of each waveguides (marked as blue line), with the wavevector mismatching $\Delta q = q - Q$ for optical modes at frequencies ${\omega_n}$ propagating inside waveguides. (c) Constructed two-dimensional synthetic lattice from (a). (d) The projected band structure of infinite synthetic lattice versus $\theta$ for $h = \kappa = g $. (e) The projected band structure with finite waveguide arrays ($m \in \left[ { - 40,40} \right]$) versus ${k_f}$ for ${\theta = \pi/2}$, which is labelled by the red dashed line in (d).
}
\end{figure}

On the other hand, the lithium niobate-on-insulator (LNOI) based integrated photonic technology has proven to be a potential platform towards manipulating light in the on-chip device
 \cite{device1,device2,device3,device4,device5,device6,device7}. In particular, a single LNOI waveguide, supporting a large second-order nonlinearity and then resulting fast electro-optic modulation
\cite{degree7qin2018prl,eom}, can be used to create the synthetic frequency dimension, where exotic physical phenomena such as frequency negative refraction and perfect imaging have been demonstrated
\cite{degree7qin2018prl}. However, following similar idea as coupling modulated rings
\cite{degree4,degree5yuan2016ol} to build a two-dimensional space including the frequency axis of light, the fabrication of multiple LNOI waveguides coupled in the spatial dimension with each one undergoing dynamic modulations is not straightforward. Under the current technology, coupling two modulated LNOI waveguides results in crosstalk of the applied external electric fields. While the experimentalist seeks the improvement of the start-of-art technology, it is of theoretical interest to study coupled LNOI waveguides with realistic fabrication possibility [see Fig.~\ref{fig1}(a)], and explore the physics in the synthetic space built in such design.

In this work, we theoretically study physical phenomena in one-dimensional weakly coupled waveguide arrays, where the electro-optic modulated waveguides and unmodulated waveguides are arranged alternatively as shown in Fig.~\ref{fig1}(a). Such a system supports a two-dimensional synthetic lattice for the travelling light along waveguides, whose dimensions are spatial and frequency ones. We choose spatial couplings between waveguides following the Su-Schrieffer-Heeger (SSH) configuration, while introduce the effective magnetic flux by adding non-uniform modulation phases in each modulated waveguide. Our results show that this hybrid system supports exotic topological edge states (\textit{quasi-edge state}s), where the wavepacket of light evolving unidirectionally along the frequency dimension is not localized at the edge. Moreover, the topological Bloch oscillation has been seen in simulations, where the oscillation of frequency modes is located at the boundary of waveguide arrays. Our work therefore not only shows important topological photonic phenomena in a hybrid synthetic lattice, but also points out potential ways to manipulate the frequency of light under topological protection in modulated waveguide arrays, which leads towards possible applications for controlling light in integrated photonics.

We start from studying a realistic experimental platform based on LNOI as shown in Fig. ~\ref{fig1}(a). One-dimensional waveguide arrays are proposed to be fabricated in micrometer-thick LNOI using proton-exchange or titanium in-diffusion method \cite{fabr1,fabr2}, as well as in lithium niobate thin films using shallow-etched or loaded waveguide array structure \cite{array1,array2,array3}. Such designed structure supports two types of waveguides (labelled by A and B respectively) weakly coupled between each others. We propose that waveguides of type A are undergoing dynamic modulations by covering positive and ground electrodes while waveguides of type B experience no modulation. The upper and lower buffering layers of SiO$_2$ (typically 2 micrometers) are introduced to reduced metallic loss from the electrodes. The bottom metal layer serves as the ground electrode. Therefore, there is no significant crosstalk of the applied electric fields between two modulated waveguides. The distance between waveguide centers of types A and B can be larger than 5 micrometers due to a weak transverse light confinement, which makes the fabrication of electrodes feasible in experiments.

In the vicinity of the reference central frequency
$\omega_0$, we assume the linear dispersion relation $\omega  = k\left( {{{\partial \omega } \mathord{\left/
 {\vphantom {{\partial \omega } {\partial k}}} \right.
 \kern-\nulldelimiterspace} {\partial k}}} \right)$.
We further consider a sinusoidal travelling-wave radio frequency modulation inside each waveguide of type A, i.e.,
$\cos \left( {\Omega t - Q{\kern 1pt} z + \phi } \right)$, where $\Omega$ is the modulation frequency,  $z$ is the propagation direction along the waveguide, $\phi$ is the modulation phase, and
$Q \approx q \equiv \Omega \left( {{{\partial k} \mathord{\left/
 {\vphantom {{\partial k} {\partial \omega }}} \right.
 \kern-\nulldelimiterspace} {\partial \omega }}} \right)$ is the modulation wavevector \cite{degree7qin2018prl,bo2qin2018pra,li2021laser}. Hence discrete modes in waveguides arrays can be excited at frequencies  [see Fig.~\ref{fig1}(b)]
\begin{equation}\label{eq1}
{\omega _n} = {\omega _0} + n\Omega ,
\end{equation}
with the effective modulation strength between the adjacent modes being $g$ and the wavevector mismatching $\Delta q=q-Q<<Q$. Note that $\Delta q = 0$ gives the phase matching condition while $\Delta q \ne 0$ leads to an effective gradient force in the synthetic frequency space \cite{degree7qin2018prl,bo2qin2018pra}. There is no direct modulation applied inside waveguides of the type B. The propagating modes at frequencies $\omega_n$ in the \textit{m}$^{\rm th}$ waveguides of type A and the (\textit{m}+1)$^{\rm th}$ waveguide of type B therefore construct a synthetic two-dimensional space labelled by $A_{m,n}$ and $B_{m+1,n}$ respectively as illustrated in Fig.~\ref{fig1}(c) where we set \textit{m} as an even integer. We further consider a non-uniform distribution of waveguides in the $x$ direction, i.e., we set the distance between centers of the \textit{m}$^{\rm th}$ waveguide of type A and the ($m+$1)$^{\rm th}$ waveguide of type B as $d_1$ and the distance between centers of the ($m-$1)$^{\rm th}$ waveguide with type B and the $m$$^{\rm th}$ waveguide of type A as $d_2$. Slightly difference between $d_1$ and $d_2$ can bring non-uniform coupling between modes at the same frequency $\omega_n$ in nearby waveguides, where the coupling strengths $\kappa$ between $A_{m,n}$ and $B_{m+1,n}$ and $h$ between waveguides $B_{m-1,n}$ and $A_{m,n}$ (which are assumed to be uniform in the spectral regime in our consideration) can be different.

Following the procedure in comparing the wave equation describing the light travelling in the waveguide along the propagation direction $z$ from Maxwell’s equations and the Schr\"{o}dinger equation describing the wavepacket dynamics with the time \cite{degree7qin2018prl,bo2qin2018pra,Eq1,Eq2}, the effective Hamiltonian describing the system can be written as
\begin{equation}\label{eq2}
H = \sum\limits_{m = {\rm{even}},n} {\left( {\kappa a_{m,n}^\dag b{}_{m + 1,n} + hb_{m - 1,n}^\dag a{}_{m,n} + ga_{m,n}^\dag a{}_{m,n + 1}{e^{ - i{\phi _m}}}} \right)}  + h.c.,
\end{equation}
where $a_{m,n}$ and $b_{m+1,n}$ ($a_{m,n}^\dag $ and $b_{m + 1,n}^\dag $) are annihilation (creation) operators for the modes $A_{m,n}$ and $B_{m+1,n}$, respectively. We assume modulations in every two nearby waveguides of type A are different by a constant phase $\theta$, which can be constructed by applying the same travelling-wave signal shapes with a constant pulse delay on each electrode. Therefore, the corresponding modulation phase on the \textit{m}$^{\rm th}$ waveguide of type A can be written as $\phi_m=m\theta/2$, as shown in Fig.~\ref{fig1}(c). The constructed synthetic two-dimensional lattice hence provides a set of standard one-dimensional SSH lattices \cite{SSH1,SSH2,SSH3,SSH4} along the spatial dimension at each mode with the same frequency, which are then connected along the synthetic frequency dimension. Moreover, the introduced distribution of modulation phases in the system creates the effective gauge potential ${|{\rm \textbf{A}}_{\rm eff}|} \propto \theta $ in the synthetic two-dimensional lattice \cite{fang2012}. Analysis of the Hamiltonian in Eq.~(\ref{eq2}) can be used to understand the system and explore the field propagating dynamics through waveguide arrays.

To start our analysis, we first consider a uniform distribution of the waveguide arrays $h=\kappa$, and also assume that $\kappa=g$. If we consider infinite modes in both $x$ and $f$ axes and choose $\theta$ so that ($2\pi/\theta$) gives an integer, the synthetic lattice holds the translational symmetry with the periodicity of $\Omega$ in the $f$ direction and the periodicity of $(2\pi /\theta ) \times ({d_1} + {d_2})$ in the $x$ direction. Following the well-established method in calculating the band structure with the (effective) magnetic flux \cite{flux1,flux2,flux3}, we obtain the projected band structure of our system, $\varepsilon$, versus $\theta$ from 0 to $2\pi$ as shown in Fig.~\ref{fig1}(d), which shows Hofstandter butterfly-like spectrum. Due to additional lattice sites on waveguides of type B in each plaquette, the spectrum exhibits double amounts of clear open gaps for most values of $\theta$, compared to the standard Hofstandter butterfly spectrum \cite{flux1,flux2}. As an example, we choose $\theta=\pi/2$ and see 4 clear gaps along the red line in Fig.~\ref{fig1}(d). To observe details, we plot the band structure at $\theta=\pi/2$ with open boundary in the $x$ direction but still assuming infinite sites along the frequency dimension. This makes $k_f$, which is reciprocal to the $f$ axis, being a good quantum number. The corresponding Hamiltonian in the $k_f$ space is
\begin{equation}\label{eq3}
\begin{array}{l}
{H_{m,{k_f}}} = \sum\limits_{m = {\rm{even}}} {\left( {\kappa a_{m,{k_f}}^\dag b{}_{m + 1,{k_f}{\kern 1pt} } + hb_{m - 1,{k_f}{\kern 1pt} }^\dag a{}_{m,{k_f}} + h.c.} \right)} \\
\quad \quad \quad \;\; + 2g\sum\limits_{m = {\rm{even}}} {a_{m,{k_f}}^\dag a{}_{m,{k_f}}\cos \left( {{k_f}{\kern 1pt} \Omega  + {\phi _m}} \right).}
\end{array}
\end{equation}
In calculations, we choose 41 waveguides of type A and 40 waveguides of type B, which makes $m \in \left[ { - 40,40} \right]$ where even (odd) integers refer to the waveguide of type A (B). The band structure versus $k_f$ is shown in Fig. 1(e). One sees there are 4 pairs of edge states inside 4 gaps, which are double compared to the conventional energy diagram in the traditional rectangular lattice \cite{degree4,degree5yuan2016ol,fang2012,tradition}. The duplicity of edge states is understandable due to pairs of sites (A and B) on the boundaries along the $x$ axis. Here we shall note that the energy of the band structure actually refers to the shift of wavevector for a probe signal propagating in the $z$ direction \cite{shift}.

\begin{figure}[htb]
\center
\includegraphics[width=16cm]{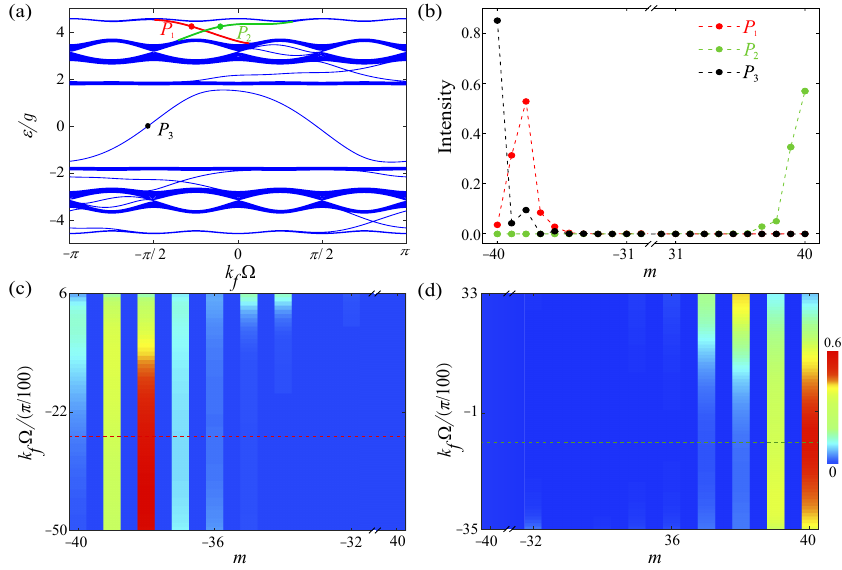}
\caption{\label{fig2}(a) The projected band structure with finite waveguide arrays ($m \in \left[ { - 40,40} \right]$), where $h=3g$, $\kappa=g$, and $\theta=\pi/2$. (b) Distributions of field intensities for states at $P_1$ (red), $P_2$ (green), and $P_3$ (black) labelled in (a), respectively. (c) and (d) Normalized intensity distributions of topological edge states labelled by red and green lines in (a) in a range of $k_f$.
}
\end{figure}

We next explore the case that $h \ne \kappa $, corresponding to connected SSH lattices under the effective magnetic flux. In particular, we set $h=3g$ while $\kappa=g$, and calculate the band structure [see Fig.~\ref{fig2}(a)] with other parameters being the same as those in Fig.~\ref{fig1}(e). Similarly, there are four pairs of topological edge states distributed in band gaps. However, the striking feature is that the middle bulk band near $\varepsilon  = 0$ in Fig.~\ref{fig1}(e) is open and there exhibits an isolated band as shown in Fig.~\ref{fig2}(a). This isolated band corresponds to the topological boundary state in the one-dimensional SSH lattice, but shows a cosine-like band shape, meaning the connection of such boundary states along the additional frequency dimension. In Fig.~\ref{fig2}(b), we plot the distribution of field intensities of the isolated band indicated by the black dot in Fig.~\ref{fig2}(a), where the energy of field is located at the left boundary and decays into the bulk. On the other hand, field intensity distributions of pairs of edge states in the top band gap labelled by red and green dots are also plotted. While for one edge state (green dot), the field is localized at the right boundary, the other state (red dot) exhibits an interesting feature that the field is focused on the third waveguide (the second waveguide of type A) from the left and decays exponentially into two sides, which refers to \textit{quasi-edge state}. This exotic quasi-edge state is a unique consequence from the effective magnetic flux and the non-trivial connectivity from the SSH configuration, which is very different from the edge states in the conventional two-dimensional square lattice under the magnetic flux \cite{degree4,degree5yuan2016ol,fang2012,tradition}. To further explore the quasi-edge state, we plot normalized intensity distributions of topological edge states labelled by red and green lines versus a range of $k_f$ in Figs.~\ref{fig2}(c) and \ref{fig2}(d), respectively. One notes that intensity distributions in Fig.~\ref{fig2}(d) show conventional characteristics  of the edge state, where energy of light is localized on the right boundary when it is in the middle of the gap, but this phenomenon fades away when it closes to the bulk bands. However, as illustrated in Fig.~\ref{fig2}(c), one sees that similar phenomenon when  $k_f\Omega$ goes towards $6\pi/100$ near the bulk bands, but in the majority of the gap, the distributions of light exhibit the maximum peak in the third waveguide from the left, indicating the quasi-edge state, which results from the competition between two topological effects, namely the quantum Hall effect and the SSH effect.

To explore exotic quasi-edge state as well as the isolated band in the synthetic lattice in Fig.~\ref{fig1}(c), we simulate the wave propagating dynamics inside waveguide arrays from the Hamiltonian (\ref{eq2}). We suppose the wave function of light as
\begin{equation}\label{eq4}
\left| {\varphi (z)} \right\rangle  = \sum\limits_{m = {\rm{even}},n} {\left( {{v_{a,\,m,{\kern 1pt} n}}a_{m,n}^\dag  + {v_{b,\,m + 1,{\kern 1pt} n}}b_{m + 1,n}^\dag } \right)} \left| 0 \right\rangle ,
\end{equation}
where $v_{a,m,n}$, $v_{b,m+1,n}$ are the amplitudes of light for photon states at modes $A_{m,n}$ and $B_{m+1,n}$, respectively. We again consider 81 waveguides, and by using the Schrödinger-like equation, $i\frac{d}{{d{\kern 1pt} z}}|\varphi (z)\rangle  = H\left| {\varphi (z)} \right\rangle $, we obtain coupled-mode equations
\begin{equation}\label{eq5}
\begin{array}{l}
i \frac{d}{{dz}}{v_{a,m,n}} = \kappa {v_{b,m + 1,n}} + h{v_{b,m - 1,n}} + g({v_{a,m,n - 1}}{e^{ - i{\phi _m}}} + {v_{a,m,n + 1}}{e^{i{\phi _m}}}) + is{\delta _{m,M}}{\delta _{n,0}},\\
 i \frac{d}{{dz}}{v_{b,m + 1,n}}{\kern 1pt}  = \kappa {v_{a,m,n}} + h{v_{a,m + 2,n}}\left( {m = {\rm even}} \right).
\end{array}
\end{equation}
Here we choose the exciting source $s = e{}^{i\Delta kz}$ where $\Delta k$ stands for the wavevector mismatching of the input light from the reference mode ${k_0} = {{{\omega _0}} \mathord{\left/
 {\vphantom {{{\omega _0}} {(\partial \omega /\partial k)}}} \right.
 \kern-\nulldelimiterspace} {(\partial \omega /\partial k)}}$. The source is injected into waveguide arrays from adding an auxiliary waveguide (on the side or beneath waveguide arrays \cite{source1,source2,source3}), very weakly coupled to the $M^{\rm th}$ waveguide with the loss from coupling between the auxiliary waveguide and waveguide arrays being negligible. To better illustrate the topological protection, we set the artificial boundary at the frequency dimension and assume $n \in [-40,40]$, which can be realized by carefully engineering the dispersion curve of the waveguide \cite{degree5yuan2016ol,shan}. However, such artificial boundary is not mandatory to be designed in order to observe exotic quasi-edge states and isolated band from the synthetic lattice.

 \begin{figure}[htb]
\center
\includegraphics[width=16cm]{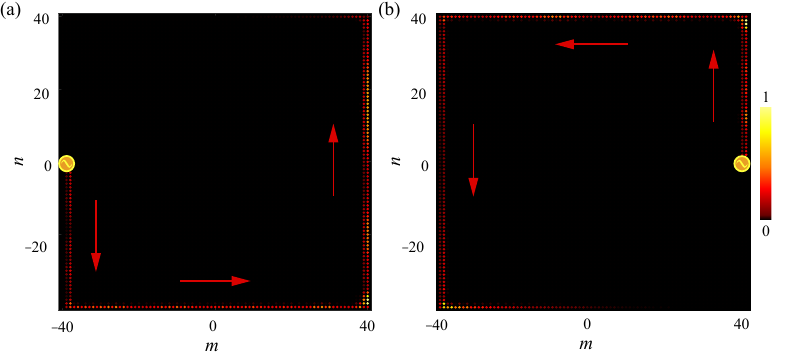}
\caption{\label{fig3} (a) and (b) Simulation results at $z=50g^{-1}$ in the synthetic space with an external source $s=e^{i\Delta kz}$ and $\Delta k=4.237g$ excited at the leftmost and rightmost waveguides, respectively.
}
\end{figure}

We next provide simulations corresponding to the (quasi-)edge state by selectively exciting the leftmost waveguide and the rightmost waveguide respectively at $\Delta k=4.237g$ near the $0^{\rm th}$ frequency mode. We assume the length of waveguide arrays being $50g^{-1}$. Figs.~\ref{fig3}(a) and \ref{fig3}(b) show the corresponding simulation results of the field distribution on different waveguides and frequency modes on the synthetic lattice at the output of waveguide arrays. One sees topologically-protected one-way propagating modes in both cases, where the field gets turned at the corner and transports in the synthetic two-dimensional space unidirectionally without back-reflection. Moreover, as indicated by band structure analysis, we find quasi-edge state near the left boundary of the synthetic lattice, where the third waveguide from the left is excited for the most. For the external excitation on the leftmost waveguide, the energy of light tunnels into the $-$38$^{\rm th}$ waveguide quickly and then gets most frequency down conversion inside the $-$38$^{\rm th}$ waveguide mostly, while for the excitation on the right, the light still transports to the $-$38$^{\rm th}$ waveguide after it circulates counter-clock-wisely through the top boundary. Exotic one-way edge state therefore exhibits where the first two waveguides from the left are very weakly excited in both cases.

\begin{figure}[htb]
\center
\includegraphics[width=16cm]{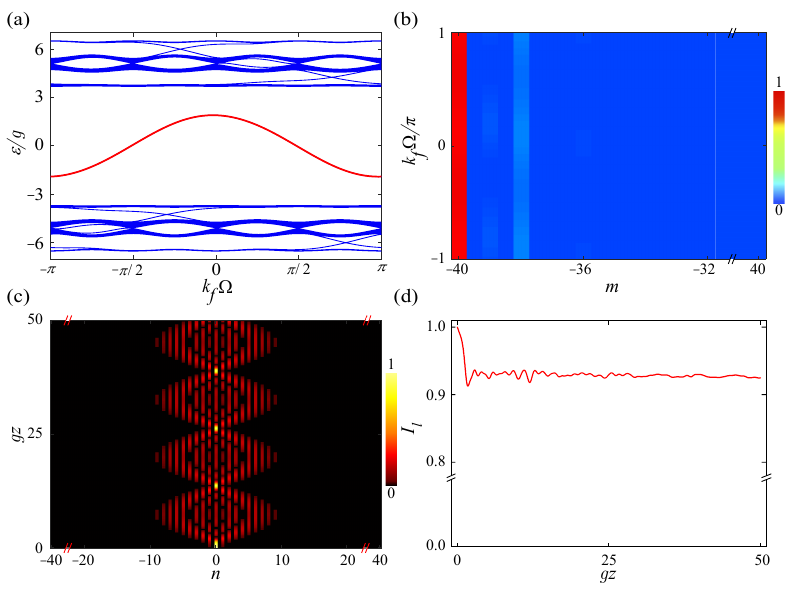}
\caption{\label{fig4} (a) The projected band structure with finite waveguide arrays ($m\in [-40,40]$), where $h=5g$, $\kappa=g$, and $\theta=\pi/2$, where middle isolated band is labelled in red. (b) Normalized intensity distributions of isolated band by the red curve in (a) for a range of $k_f$. (c) The evolution of the field propagating along the $z$ direction distributed on the $n$ frequency modes inside the leftmost waveguide. (d) The ratio of energy distribution on the leftmost waveguide $I_l$ versus the propagation distance $z$.
}
\end{figure}

Previously we showed the isolated band near $\varepsilon = 0$ in Fig.~\ref{fig2}(a). Further increasing $h$ can enlarge the middle gap and the isolated band gives a better clean cosine-shape, which is shown in Fig.~\ref{fig4}(a) with $h=5g$ and $\kappa=g$. In Fig.~\ref{fig4}(b), we plot normalized intensity distributions of the isolated band versus $k_f$ in the whole range of the first Brillouin zone, i.e., $k_f$ in $[-\pi/\Omega,\pi/\Omega]$. We can see that the field mostly distributes on the left boundary due to the SSH configuration in the synthetic lattice. Hence the cosine-shape isolated band provides a unique topological platform to explore topological Bloch oscillation along the synthetic frequency dimension.

Based on the effective Hamiltonian in Eq.~(\ref{eq2}), we consider the wavevector mismatching case with $\Delta q \ne 0$, which leads to the Hamiltonian:
\begin{equation}\label{eq6}
\tilde H = \sum\limits_{m = {\rm even},n} {\left[ {\kappa a_{m,n}^\dag b{}_{m + 1,n} + hb_{m - 1,n}^\dag a{}_{m,n} + ga_{m,n - 1}^\dag a{}_{m,n}{e^{ - i(\Delta qz + {\phi _m})}}} \right]}  + h.c.,
\end{equation}
implying a constant effective force $F=-\Delta q$ \cite{frequency3li2021sci,bo1yuan2016optica,bo3,force}.

In simulations, we choose a Gaussian-shape pulse as the excitation source, which reads as
\begin{equation}\label{eq7}
s(z) = {e^{ - 8\ln 2{{(gz - 1)}^2}}},
\end{equation}
and excite the 0$^{\rm th}$ mode in the leftmost waveguide. In Fig.~\ref{fig4}(c), we show the evolution of the field distributed on frequency modes inside the leftmost waveguide. The Bloch oscillation feature in the frequency dimension with the spatial periodicity about ${Z_B} = 2\pi /\left| {\Delta q} \right| \approx 12.6{g^{ - 1}}$ can be seen. In order to explore the topological feature of the Bloch oscillation, we define the ratio of energy distribution on the leftmost waveguide ${I_l} = \sum\limits_n {{{\left| {{v_{a, - 40,n}}} \right|}^2}} /\sum\limits_{m = {\rm{even}},n} {\left( {{{\left| {{v_{a,m,n}}} \right|}^2} + {{\left| {{v_{b,m + 1,n}}} \right|}^2}} \right)} ,$ We plot $I_l$ versus the propagation distance $z$ in Fig.~\ref{fig4}(d). One sees that $I_l$ slightly drops from 1 and then quickly stabilizes near $93\%$ throughout the entire propagation in waveguide arrays. The energy of light is mainly localized at the leftmost waveguide during the evolution of the Bloch oscillation, indicating the nature of the topological boundary state in the system.

In summary, we propose LNOI waveguide arrays with the SSH configuration undergoing dynamic modulations, which supports a two-dimensional hybrid lattice under the effective gauge field in a synthetic space including spatial and frequency dimensions. The combination of physical consequences from the effective magnetic field and the SSH configuration brings quasi-edge state, where the topological one-way mode propagates in the vicinity of the boundary. We find a cosine-shape isolated band structure, which supports the topological Bloch oscillation along the frequency dimension localized at the spatial boundary. Our proposed model is highly relevant to the current start-of-art LNOI technology, which could be used for further experimental achievements pointing to on-chip applications. Moreover, this model is also valid in other platforms including static waveguide arrays \cite{degree8,Eq2,array}, ring resonators \cite{ring}, temporal modulated photonic crystals \cite{fang2012}, time-multiplexed network \cite{network1,network2}, and cold atoms \cite{cold1,cold2}, which shows an important way to explore topological phenomena in both real space and synthetic space.

The research is supported by National Natural Science Foundation of China (12122407, 11974245, and 12104297), National Key R\&D Program of China (2017YFA0303701), Shanghai Municipal Science and Technology Major Project (2019SHZDZX01), and Natural Science Foundation of Shanghai (19ZR1475700). L.Y. acknowledges the support from the Program for Professor of Special Appointment (Eastern Scholar) at Shanghai Institutions of Higher Learning. X.C. also acknowledges the support from Shandong Quancheng Scholarship (00242019024).

\bibliography{References}% Produces the bibliography via BibTeX.

\end{document}